\documentstyle[12pt,aaspp4]{article}

\begin{document}

%
\def\n{\footnotemark}
\def\IUE{{\it IUE}}
\def\HST{{\it HST}}
\def\deg{$^{\rm o}$}
\def\degC{$^{\rm o}$C}
\def\arcsec{\ifmmode '' \else $''$\fi}
\def\arcmin{$'$}
\def\arcsecpoint{\ifmmode ''\!. \else $''\!.$\fi}
\def\arcminpoint{$'\!.$}
\def\kms{\ifmmode {\rm km\ s}^{-1} \else km s$^{-1}$\fi}
\def\Msun{\ifmmode {\rm M}_{\odot} \else M$_{\odot}$\fi}
\def\Lsun{\ifmmode {\rm L}_{\odot} \else L$_{\odot}$\fi}
\def\Zsun{\ifmmode {\rm Z}_{\odot} \else Z$_{\odot}$\fi}
\def\ergsAcm{ergs\,s$^{-1}$\,cm$^{-2}$\,\AA$^{-1}$}
\def\ergscm2{ergs\,s$^{-1}$\,cm$^{-2}$}
\def\qo{\ifmmode q_{\rm o} \else $q_{\rm o}$\fi}
\def\Ho{\ifmmode H_{\rm o} \else $H_{\rm o}$\fi}
\def\ho{\ifmmode h_{\rm o} \else $h_{\rm o}$\fi}
\def\ltsim{\raisebox{-.5ex}{$\;\stackrel{<}{\sim}\;$}}
\def\gtsim{\raisebox{-.5ex}{$\;\stackrel{>}{\sim}\;$}}
\def\vFWHM{\ifmmode v_{\mbox{\tiny FWHM}} \else
            $v_{\mbox{\tiny FWHM}}$\fi}
\def\CCF{\ifmmode F_{\it CCF} \else $F_{\it CCF}$\fi}
\def\ACF{\ifmmode F_{\it ACF} \else $F_{\it ACF}$\fi}
\def\Halpha{\ifmmode {\rm H}\alpha \else H$\alpha$\fi}
\def\Hbeta{\ifmmode {\rm H}\beta \else H$\beta$\fi}
\def\Hgamma{\ifmmode {\rm H}\gamma \else H$\gamma$\fi}
\def\Hdelta{\ifmmode {\rm H}\delta \else H$\delta$\fi}
\def\Lya{\ifmmode {\rm Ly}\alpha \else Ly$\alpha$\fi}
\def\Lyb{\ifmmode {\rm Ly}\beta \else Ly$\beta$\fi}
\def\Lyg{\ifmmode {\rm Ly}\beta \else Ly$\gamma$\fi}
\def\hi{H\,{\sc i}}
\def\hii{H\,{\sc ii}}
\def\hei{He\,{\sc i}}
\def\heii{He\,{\sc ii}}
\def\ci{C\,{\sc i}}
\def\cii{C\,{\sc ii}}
\def\ciii{\ifmmode {\rm C}\,{\sc iii} \else C\,{\sc iii}\fi}
\def\civ{\ifmmode {\rm C}\,{\sc iv} \else C\,{\sc iv}\fi}
\def\ni{N\,{\sc i}}
\def\nii{N\,{\sc ii}}
\def\niii{N\,{\sc iii}}
\def\niv{N\,{\sc iv}}
\def\nv{N\,{\sc v}}
\def\oi{O\,{\sc i}}
\def\oii{O\,{\sc ii}}
\def\oiii{O\,{\sc iii}}
\def\o5007{[O\,{\sc iii}]\,$\lambda5007$}
\def\oiv{O\,{\sc iv}}
\def\ov{O\,{\sc v}}
\def\ovi{O\,{\sc vi}}
\def\neiii{Ne\,{\sc iii}}
\def\nev{Ne\,{\sc v}}
\def\neviii{Ne\,{\sc viii}}
\def\mgi{Mg\,{\sc i}}
\def\mgii{Mg\,{\sc ii}}
\def\siIV{Si\,{\sc iv}}
\def\siIII{Si\,{\sc iii}}
\def\siII{Si\,{\sc ii}}
\def\si{S\,{\sc i}}
\def\sii{S\,{\sc ii}}
\def\siii{S\,{\sc iii}}
\def\siv{S\,{\sc iv}}
\def\sv{S\,{\sc v}}
\def\svi{S\,{\sc vi}}
\def\caii{Ca\,{\sc ii}}
\def\feii{Fe\,{\sc ii}}
\def\feiii{Fe\,{\sc iii}}
\def\alii{Al\,{\sc ii}}
\def\aliii{Al\,{\sc iii}}
\def\piv{P\,{\sc iv}}
\def\pv{P\,{\sc v}}
\def\cliv{Cl\,{\sc iv}}
\def\clv{Cl\,{\sc v}}
\def\o{\o}
%

\title{Locally Optimally Emitting Clouds and \\
the Origin of Quasar Emission Lines}

\author{Jack Baldwin}
\affil{Cerro Tololo Interamerican Observatory\altaffilmark{1}, Casilla
603, La Serena, Chile}
\author{Gary Ferland, Kirk Korista, and Dima Verner}
\affil{Department of Physics \& Astronomy, University of Kentucky,
Lexington, KY 40506}

\altaffiltext{1}{Operated by the Association of Universities for
Research in Astronomy Inc.\ (AURA) under cooperative agreement with the
National Science Foundation}

\begin{abstract}
The similarity of quasar line spectra has been taken as an indication
that the emission line clouds have preferred parameters, suggesting
that the environment is subject to a fine tuning process.   We show
here that the observed spectrum is a natural consequence of powerful
selection effects.   We computed a large grid of photoionization models
covering the widest possible range of cloud gas density and distance
from the central continuum source.  For each line only a narrow range
of density and distance from the continuum source results in maximum
reprocessing efficiency, corresponding to ``locally optimally emitting
clouds'' (LOC).  These parameters depend on the ionization and
excitation potentials of the line, and its thermalization density.  The
mean QSO line spectrum can be reproduced by simply adding together the
full family of clouds, with an appropriate covering fraction
distribution.  The observed quasar spectrum is a natural consequence of
the ability of various clouds to reprocess the underlying continuum,
and can arise in a chaotic environment with no preferred pressure, gas
density, or ionization parameter.
\end{abstract}

\keywords{quasars: emission lines}

\newpage

\section{Introduction}

The question of the origin of quasar emission line regions is central
to calibrating these objects as probes of the Universe at the epoch of
galaxy formation.  Once we understand the spectrum in detail there is
the potential of measuring their distances (and hence \qo\/ and \Ho\/)
and of determining the composition of the gas, thus constraining
nucleosynthesis during galaxy formation.

Although the promise is great, progress has been slow.  The lines are
produced by a dilute gas far from LTE.  Its spectrum is sensitive to
line formation details, particularly the cloud gas density and flux of
radiation (Netzer 1992).  These are often combined into a single
``ionization parameter'', the ratio of photon to gas densities.  The
earliest observations showed that the family of quasars have
surprisingly similar emission line spectra (Davidson \& Netzer 1979),
and this is often taken as evidence that a process adjusts clouds to a
certain ionization parameter (Davidson 1977, Krolik, McKee, \& Tarter
1981).  The possible presence of an unspecified tuning mechanism made
it impossible to know how to interpret correlations between emission
line parameters and other properties ({\em e.g.}, Baldwin, Wampler, \&
Gaskell 1989).  Finely tuned cloud parameters suggest a highly ordered
environment, not the chaotic one that might naturally result during the
initial stages of galaxy formation.

More recent work by Rees, Netzer, \& Ferland (1989), Krolik et
al.\ (1991), Goad, O'Brien, \& Gondhalekar (1993), and O'Brien, Goad,
\& Gondhalekar (1994, 1995) have investigated distributed emission line
cloud properties, but still assumed the existence of a single cloud
pressure law with radius. Here we use a much more general approach to
show that a powerful selection effect operates, so that a random mix of
cloud properties will produce the spectrum of a typical QSO.  There is
no spectroscopic evidence for finely tuned cloud parameters.  The
spectrum of a typical quasar is consistent with spectral formation in
an environment with no preferred pressure, gas density, or ionization
parameter.

\section{A Single Cloud}

We model individual clouds using the following assumptions: a single
column density of N(H)~$= 10^{23}$~cm$^{-2}$, constant hydrogen density
throughout each cloud, and solar metallicity. Although below we show
results for a single cloud column density, we expect there to be a
broad range of column densities present in the BLR. Preliminary
investigations into the sensitivity of the results to the presence of a
range of column densities indicate that they are not strongly affected
as long as optically thin clouds do not dominate the total line
emission. With the exception of \ion{N}{5} $\lambda$1240, most of the
prominent emission lines are not strongly affected by our choice of
metallicity (see Hamann \& Ferland 1993).

The ionizing continuum shape chosen was a combination of a $f_{\nu}
\propto \nu^{-0.3} exp(-h\nu/kT_{cut}$) UV-bump with an X-ray power
law of the form $f_{\nu} \propto \nu^{-1}$ spanning 13.6~eV to
100~keV.  The UV-bump cutoff temperature, $T_{cut}$, was chosen such
that the energy in the UV-bump peaked at 48~eV. The two continuum
components were combined using a typical QSO UV to X-ray logrithmic
spectral slope of $\alpha_{ox} = -1.4$ (note the explicit minus sign).
Korista et al.\ (1996) will explore in detail the dependencies of the
emitted line spectrum on continuum shape, chemical abundances, and
cloud column density.

The remaining parameters are the cloud particle density $n(H)$
(cm$^{-3}$) and the photon flux of H-ionizing radiation $\Phi(H)$
(cm$^{-2}$~s$^{-1}$). For an isotropic continuum source $\Phi(H)$ is a
direct measure of the distance between the clouds and the continuum
source $r$. The variables $r$ and $n(H)$ directly describe the overall
structure of the BLR, and we do not combine them into an ionization
parameter $U(H) = \Phi(H)/n(H)c$.

\section{Locally Optimally-emitting Clouds (LOC)}

Figure~1 shows the results of a large series of photoionization
calculations in which $\Phi(H)$ and $n(H)$ were varied.  Lines are
presented as equivalent widths for full geometrical coverage, referred
to the incident continuum at 1215~\AA\/.  This equivalent width is a
direct measure of the cloud's reprocessing efficiency.

Collisionally excited lines such as \ion{C}{4} $\lambda$1549 show a
band of efficient reprocessing running at constant $U(H)$ along a
diagonal ridge from high $\Phi(H)$ and $n(H)$, to low $\Phi(H)$ and
$n(H)$.  Along this ridge, corresponding to $\log~U(H) \approx -1.5$
for $\lambda$1549, the gentle decrease in the $W_{\lambda}$ at the high
densities is caused by line thermalization --- high density clouds are
continuum sources.  The line equivalent width decreases sharply when
moving orthogonal to the ridge because $U(H)$ is either too low (lower
right) or too high (upper left) to produce the line efficiently. Clouds
with $\log~U(H) \gtsim 0.2$ are optically thin so reprocess little of
the incident continuum. \ion{C}{4} can be contrasted with a low
ionization line such as \ion{Mg}{2} $\lambda$2798, whose peak
$W_{\lambda}$ is shifted to lower $U(H)$, or to the high ionization
\ion{O}{6} $\lambda$1034, shifted to higher $U(H)$ ($\approx 1$). Note
that the contours in Figure~1 are logarithmic; in linear space they are
sharply peaked. We then see that the line formation radius naturally
depends on the ionization potential (IP), modulated by thermalization
density; low IP lines form at large radii and high IP at small radii, a
result consistent with line-continuum reverberation measurements
(Peterson 1993).

Recombination lines of H$^o$ and He$^+$ emit over a wider area on the
density--flux plane, including the low $\Phi(H)$ -- high $n(H)$ regions
since both ions still exist under these conditions. On the whole,
\Lya\/ forms closer in than \Hbeta\/. This is because \Lya\/ is a
resonance line and is collisionally suppressed at high densities, while
\Hbeta\/ is suppressed at high flux due to larger excited state optical
depths (Ferland, Netzer, \& Shields 1979). This will make \Lya\/
respond first to continuum variations and make its line profile broader
in the wings if cloud motions are purely virial. The first effect is
observed (Clavel et al.\ 1991; Peterson et al.\ 1991); the second
effect is often but not always observed (Zheng 1992; Netzer et
al.\ 1995). For \ion{He}{2} $\lambda$1640, the region of most efficient
reprocessing is a diagonal plateau (rather than a ridge), peaking and
saturating at high density and flux, but also diminishing at very low
ionization parameter. This peak and much of the plateau are at
significantly larger flux than for \Hbeta\/, thus $\lambda$1640 is
formed in gas lying closer to the continuum source.  This is also
consistent with the reverberation measurements (Korista et al.\ 1995).

\section{The Integrated LOC Spectrum}

The equivalent widths for full coverage, as plotted in Figure~1,
represent the efficiency with which the individual LOC clouds can
reprocess ionizing continuum radiation into line radiation. These
reprocessing efficiencies act as a filter whose output is the QSO's
emission line spectrum.  Each line can be formed efficiently only at
a particular location in the density--flux plane.  This location is
different for different lines.  If the individual clouds are
distributed over a wide range on the density--flux plane, then most
emission lines will be produced with high efficiency, so long as that
component has a significant covering fraction.

The simplest description of this type of system is to take, for each
separate line, the extreme case of considering only emission from a
cloud with the optimal density and flux parameters for the line in
question. Under these conditions the relative intensity of each
emission line would be directly proportional to the maximum
reprocessing efficiency, for uniform covering fraction. Column~2 in
Table~1 presents a mean observed QSO spectrum, and in column~3 are the
results of this simple model. The two are in good agreement.

A more realistic refinement is to assume some distribution of clouds on
the density--flux plane and use that as a weighting function to
integrate over the distribution of reprocessing efficiencies. The
limits of this integration are set by physical or observational
considerations: clouds at large distances from the continuum source
($\log \Phi(H) < $18) will form graphite grains (Sanders et al.\ 1989)
and therefore have very low emissivity (Netzer \& Laor 1993); very low
density clouds ($n(H) < 10^8$ cm$^{-3}$) are ruled out by the absence
of broad forbidden lines, such as [\ion{O}{3}] $\lambda$4363. Clouds at
very small radii or with very high densities are either at the Compton
temperature or are so dense that they are continuum rather than
emission line sources. Thus to produce the emission line spectrum, one
simply integrates the line emission from clouds along each axis in
Figure~1 ($\log n(H) \geq 8$, $\log \Phi(H) \geq 18$).

Remaining to be specified are the distribution of clouds as functions
of distance from the continuum source and gas density, appearing in the
definition of the emission line luminosity, \begin{equation} L_{line}
\propto \int\!\!\int r^{2}F(r)\,f(r)\,g(n)\,dn\,dr, \end{equation}
where F(r) is the emission line flux of a single cloud at radius $r$,
and $f(r)$ and $g(n)$ are the cloud covering fractions with radius and
gas density, respectively. Column~4 of Table~1 shows the results for
the simple case of $f(r) \propto constant$ and $g(n) \propto n^{-1}$;
the results are sensitive to these distributions and will be discussed
in Korista et al.\ (1996).  The agreement with the observations again
is quite good, with the exception of \ion{N}{5} $\lambda$1240
(underpredicted, as usual for solar abundances, see Hamann and Ferland
1993). Some of the other weaker lines in Table~1 whose predicted
relative intensities lie near the bottom of the observed distribution
should also be larger in gas of higher metallicity expected in
QSOs ({\em e.g.}, \ion{Si}{3}], \ion{Si}{4}, \ion{Al}{3}). The
\Lya\//\Hbeta\/ ratio is still problematical, with large uncertainties
in intrinsic reddening and radiative transfer in the Balmer lines
(Netzer et al.\ 1995). In this example, a total geometrical covering
fraction of 18\% will produce the typical $W_{\lambda}$(\Lya\/) $=$
100~\AA\/.

\section{Conclusions}

These results show that line emission from the BLR is dominated by
powerful selection effects. Individual BLR clouds can be thought of as
just machines for reprocessing radiation. As long as there are enough
clouds at the correct radius and with the correct gas density to
efficiently form a given line, the line will be formed with a relative
strength which turns out to be very similar to the one actually
observed. Thus there is no spectroscopic evidence for preferred cloud
parameters. Observed QSO spectra carry information mostly about global
properties such as the continuum shape, the clouds' overall
metallicity, and their distribution, rather than about the details of
individual clouds.

This in turn shows that a jumbled or chaotic cloud environment can be
the source of the lines.  The spectrum of a typical quasar may be
produced in a scrambled environment similar to that which might occur
in the center of a proto-galaxy, with rapid star formation, evolution,
and supernova activity.

A ramification of the fact that lines form at different places is that
we can qualitatively reproduce many of the observed variability and
line width differences.  This picture also provides a natural
explanation for why the sizes of the emission line regions scale with
luminosity (Peterson 1993); lines naturally form at the appropriate
ionizing fluxes, and the corresponding radii scale with luminosity.

Many aspects of the LOC picture are under investigation: sensitivity of
results to ionizing continuum shape, metallicity, cloud covering factor
distributions with radius and gas density, resulting line profiles and
reverberation. The likely presence of a broad range of cloud column
densities converts the LOC plane to a cube ($\Phi(H), n(H), N(H)$).

If the LOC picture proves to be correct then our understanding of the
BLR will have undergone the same change as did our view of the Narrow
Line Region when it was realized that all of the forbidden lines were
radiating near their critical densities (Filippenko \& Halpern 1984).
Simple averaging, not a hidden hand or unknown physics, is at work.

We are grateful to Mark Bottorff, Moshe Elitzur, and the referee Simon
Morris for helpful comments. This work was supported by the NSF (AST
93-19034), NASA (NAGW-3315), and STScI grant GO-2306.
%
%
%
\begin{table}[h]
\begin{center}
\begin{tabular}{cccc}
\multicolumn{4}{c}{\sc TABLE~1}
\\[0.2cm]
\multicolumn{4}{c}{\sc Observed and Predicted Line Intensities}
\\[0.2cm]
\hline
\hline
\\[0.01cm]
\multicolumn{1}{c}{Emission Line}
&\multicolumn{1}{c}{Observed Intensity$^a$}
&\multicolumn{1}{c}{Maximum Reprocessing}
&\multicolumn{1}{c}{LOC Integration$^b$} \\
\multicolumn{1}{c}{(1)} & \multicolumn{1}{c}{(2)} &
\multicolumn{1}{c}{(3)} & \multicolumn{1}{c}{(4)}
\\[0.05cm]
\hline
\ion{O}{6} $\lambda$1034$+$\Lyb\/$\lambda$1026 & 0.1--0.3 & 0.28 & 0.16 \\
\Lya\/ $\lambda$1216 & 1.00 & 1.00 & 1.00 \\
\ion{N}{5} $\lambda$1240 & 0.1--0.3 & 0.06 & 0.04 \\
\ion{Si}{4} $\lambda$1397$+$\ion{O}{4}] $\lambda$1402 & 0.08--0.24 & 0.08 &
0.06 \\
\ion{C}{4} $\lambda$1549 & 0.4--0.6 & 0.54 & 0.57 \\
\ion{He}{2} $\lambda$1640 $+$ \ion{O}{3}] $\lambda$1666 & 0.09--0.2 & 0.11 &
0.14 \\
\ion{C}{3}]$+$\ion{Si}{3}]$+$\ion{Al}{3} $\lambda$1900 & 0.15--0.3 & 0.28 &
0.12 \\
\ion{Mg}{2} $\lambda$2798 & 0.15--0.3 & 0.38 & 0.34 \\
\Hbeta\/ $\lambda$4861 & 0.07--0.2 & 0.08 & 0.09
\\[0.01cm]
\hline
\end{tabular}
\end{center}
\rm
\footnotesize
\leftskip=2.em
$^a$Intensity relative to \Lya\/ $\lambda$1216, combining data from
Baldwin, Wampler, \& Gaskell (1989), Boyle (1990), Cristiani \& Vio
(1990), Francis et al.\ (1991), Laor et al.\ 1995, Netzer et
al.\ (1995), and Weymann et al.\ (1991). \\
$^b$Co-addition of emission from clouds as described in the text.
\end{table}
%
%

%
%
%
\begin{figure}
\plotfiddle{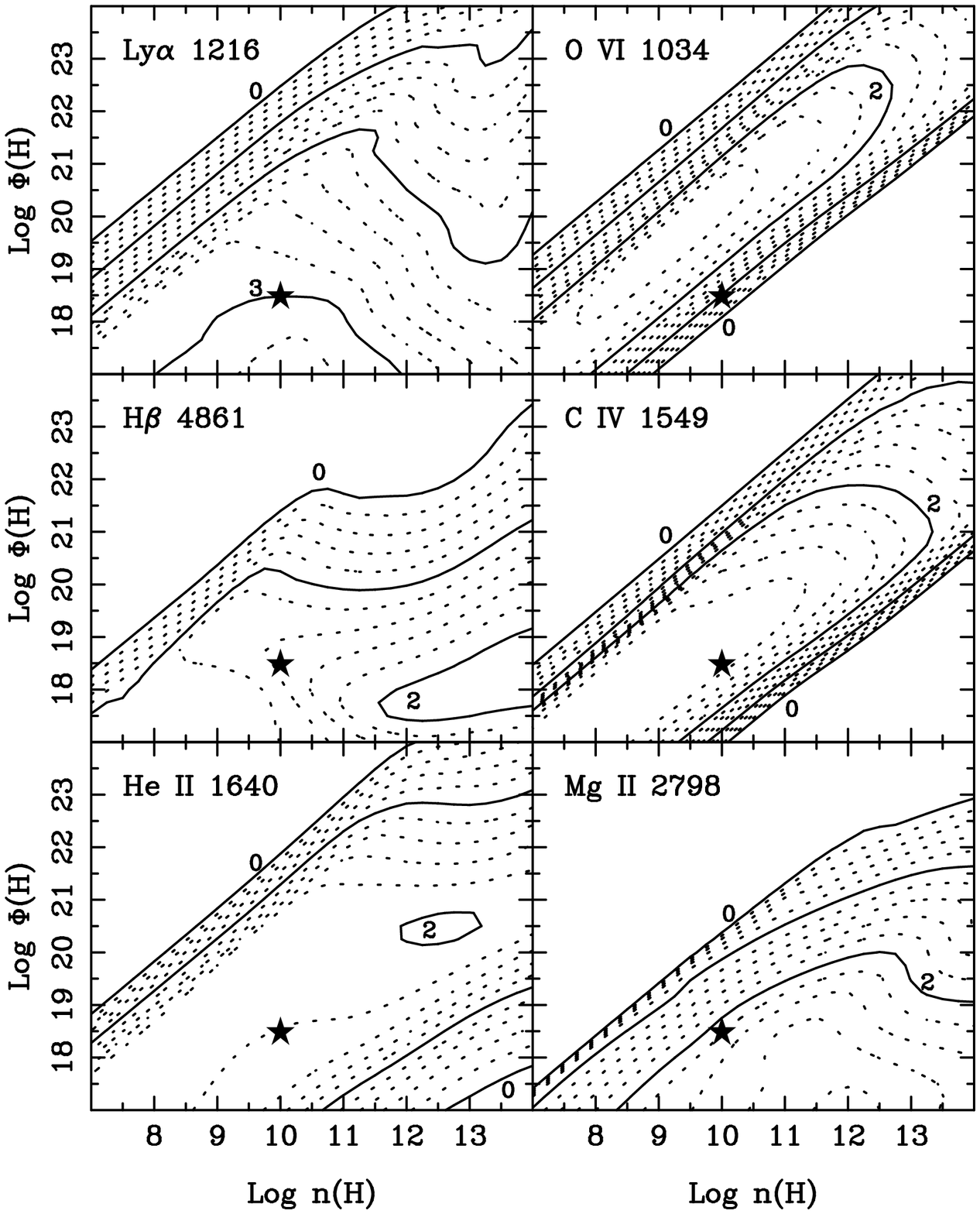}{5.0in}{0}{70}{70}{-205}{-110}
\caption{Cloud reprocessing efficiency is shown as a function of the
hydrogen density and flux of ionizing photons.  Line strengths are
expressed as equivalent widths referenced to the incident continuum at
1215~\AA\/.  Each bold line is 1 dex, and dotted lines represent 0.2
dex steps.  The smallest (1~\AA\/) and largest decade contours are
labeled for each line; note that the contour values continue to
increase beyond the largest labeled decade in all cases but \Hbeta\/
and \heii\/. The star is a reference point marking the ``standard BLR''
parameters discussed by Davidson and Netzer (1979).}
\end{figure}
\end{document}